\documentclass[trackchanges]{aastex701}

\newcommand{\rbirth}{\mbox{$R_{\rm birth}$}}

\newcommand{\feh}{\mbox{$\rm [Fe/H]$}}
\newcommand{\alphafe}{\mbox{$\rm [\alpha/Fe]$}}

\begin{document}

\title{The Absolute Age of the Open Cluster NGC 6791 and Its Implications for Galactic Archaeology and Asteroseismic Calibration}

\author[orcid=0000-0001-8706-3642]{George Dufresne}
\affiliation{Department of Physics and Astronomy, \\
Dartmouth College, \\
Hanover, NH 03755, USA}
\email{george.e.dufresne.gr@dartmouth.edu}

\author[orcid=0000-0003-3096-4161]{Brian Chaboyer}
\affiliation{Department of Physics and Astronomy, \\
Dartmouth College, \\
Hanover, NH 03755, USA}
\email{brian.c.chaboyer@dartmouth.edu}

\author[orcid=0000-0001-7337-5936]{Rayna Rampalli}
\affiliation{Department of Physics and Astronomy, \\
Dartmouth College, \\
Hanover, NH 03755, USA}
\email{rayna.rampalli.gr@dartmouth.edu}

\begin{abstract}

We present a new absolute age determination for NGC\,6791, one of the Milky Way's oldest and most metal-rich open clusters. Its unusual properties make it an important probe of inner-disk evolution and asteroseismic calibration, but its age has remained difficult to determine because of coupled uncertainties in reddening, distance, photometry, and stellar-model physics. \emph{Gaia} DR3 photometry together with detached eclipsing binaries (DEBs) in NGC\,6791 are combined with 10,000 Monte Carlo isochrone sets (marginalizing over uncertainties in composition, convective mixing processes, opacities, diffusion, nuclear reaction-rates, distance modulus, and reddening) to determine the age of NGC\, 6791. For each isochrone we build a synthetic color-magnitude diagram (CMD) that matches the observed star count in the MSTO and subgiant-branch window and injects empirical photometric scatter perpendicular to the ridgeline, enabling CMD comparisons without artificial-star tests. We assess CMD morphology using a bootstrap-calibrated two-dimensional Kolmogorov-Smirnov statistic, and add an external check based on the nearest-point metric: a coeval DEB statistic in $(M,L)$ space. These statistics are mapped to probability-density weights via bootstrap-resampling and combined into a single isochrone weight. NGC\,6791 is determined to have an age of $8.46\pm0.66$ Gyr, $[\mathrm{Fe/H}]=+0.280\pm0.079$, $Y=0.2968\pm0.0158$, $(m{-}M)_V=13.333\pm0.058$, and $E(B{-}V)=0.183\pm0.024$. Our error budget shows no single dominant contributor, and highlights differences between open-cluster and globular-cluster age errors. Combined with its super-solar metallicity, our age estimate favors an inner-Galaxy origin for NGC\,6791 and subsequent outward migration, provides a benchmark for asteroseismic calibration at high metallicity, and extends the absolute cluster age--metallicity relation to an old, metal-rich open cluster.

\end{abstract}

\keywords{
\uat{Open star clusters}{1160} --- 
\uat{Stellar evolution}{1599} --- 
\uat{Stellar evolutionary models}{2046} --- 
\uat{Eclipsing binary stars}{444} --- 
\uat{Hertzsprung-Russell diagram}{725} --- 
\uat{Gaia mission}{2360} --- 
\uat{Interstellar reddening}{853} ---
\uat{Galactic Archaeology}{2178}
}

\section{Introduction} \label{sec:intro}

Open clusters are gravitationally bound, coeval groups of stars that formed from the same giant molecular cloud and therefore share broadly similar initial chemical compositions. Old Galactic open clusters are valuable tracers of disk formation and evolution, while individual clusters provide controlled testbeds for stellar evolution by constraining chemical enrichment processes and the dynamical state of the cluster \citep{Kalirai2010,Friel1995}. \citet{Sinha2024} analyzed 26 Milky Way open clusters and found chemical homogeneity at the $3\sigma$ level, with intrinsic abundance dispersions of $\lesssim 0.02$~dex for the majority of elements.

NGC\,6791 is among the most unique  open clusters in the Milky Way, given its massive, old, and metal-rich nature. With a present-day mass of $\sim 3.6$--$5.0\times10^{3}\,M_{\odot}$ and $\sim 4{,}000$--$4{,}800$ identified members \citep{Platais2011,Ahmed2025}, it has super-solar metallicity, $[\mathrm{Fe/H}] \approx +0.31$, enhanced $[\alpha/\mathrm{Fe}]$, and an age of $\sim 8$~Gyr, with spectroscopic studies indicating little to no intrinsic metallicity spread \citep{Carraro2006,Villanova2018,Brogaard2012, Linden2017}. Its present-day location and orbit are unusual for an open cluster with these properties. Most open clusters are younger, closer to the Galactic plane, and less chemically enriched, whereas NGC\,6791 lies at a Galactocentric radius of $\sim 8$~kpc and reaches nearly $\sim 1$~kpc above the plane. Orbital constraints further indicate a pericenter of only $\sim 3$--$4.7$~kpc and an apocenter of $\sim 8.5$--$10$~kpc \citep{Bedin2006,Ahmed2025}, suggesting that its current position is difficult to reconcile with in situ formation at the solar circle. For its old age and super-solar metallicity, NGC\,6791 would more naturally be associated with the chemically evolved inner Galaxy, motivating scenarios in which it formed in the bulge or inner disk and later migrated outward. In addition, its abundance pattern places it near the transition between the high-$\alpha$ and low-$\alpha$ disk sequences: for such an old cluster, its $[\alpha/\mathrm{Fe}]$ is relatively low, while its high metallicity places it at the metal-rich end of the high-$\alpha$ population. As a result, NGC\,6791 provides an empirical constraint on the late stages of high-$\alpha$ star formation and chemical enrichment in the inner Milky Way; we return to this point in \ref{subsec:galactic_archeology}. Dynamical studies further suggest substantial historical mass loss, potentially implying an initially much larger cluster mass and highlighting NGC\,6791 as a useful link between typical open clusters and low-mass globular clusters \citep{Dalessandro2015}.

Because open clusters are approximately chemically homogeneous, fitting isochrones to their color–magnitude diagrams is a standard age-determination method, with the main-sequence turnoff (MSTO) providing the strongest age constraint \citep{Chaboyer1995}. A stellar isochrone is a tool for stellar-population analysis that traces stars of a single age across a range of masses, constructed by interpolating stellar-evolution tracks \citep{Dotter2016}. For NGC\,6791, absolute age estimates have been limited by coupled uncertainties in distance and reddening, photometric systematics and completeness, and stellar-model physics that shifts the MSTO and subgiant-branch morphology (e.g., diffusion, mixing length, overshoot, opacities, and key reaction rates).

\citet{VandenBerg_2014} analyzed NGC\,6791 using isochrone techniques and derived $[\mathrm{Fe/H}] = +0.35$, an initial helium abundance of $Y_{0} = 0.30$, $E(B-V) = 0.16$, an age of $8.0$~Gyr, and $(m-M)_V = 13.05$. A variety of complementary approaches have also been used to estimate the cluster age. From detached eclipsing binaries (DEBs), \citet{Brogaard2012} obtained $8.3 \pm 0.3$~Gyr with $Y = 0.30 \pm 0.01$, while detailed asteroseismic modelling by \citet{McKeever2019} yielded a consistent age of $8.2 \pm 0.3$~Gyr and $Y_{0} = 0.297 \pm 0.003$. White-dwarf cooling studies by \citet{Bedin2008WD} suggested a substantially younger age of $\sim 4$~Gyr, although the same study found $\sim 8$~Gyr from main-sequence turnoff fitting. Earlier isochrone work by \citet{Chaboyer1999} derived $8.0 \pm 0.5$~Gyr, with $E(B{-}V)=0.10$ and $(m{-}M)_V=13.42$. Taken together, these studies generally place NGC\,6791 at $\sim 8$~Gyr, but they also illustrate the remaining limitations in the absolute age scale. In particular, age estimates remain sensitive to coupled uncertainties in reddening and distance modulus, to assumptions about stellar input physics, and to method-specific systematics. DEB and asteroseismic constraints are powerful, but they rely on a small number of stars and still inherit stellar-model dependencies. White-dwarf cooling results have historically shown tension with turnoff-based ages. More broadly, recent work by \citet{Tayar2025} has shown that asteroseismic ages for individual cluster members can exhibit substantial dispersion and can differ from benchmark cluster ages inferred from isochrone-based analyses. These issues motivate an updated absolute age determination for NGC\,6791 that combines multiple empirical constraints within a single statistical framework.

Although these studies provide broadly consistent ages, previous determinations did not fully marginalize over the wide range of uncertainties in stellar evolution models and isochrone construction, so their quoted error bars likely do not capture the full uncertainty in the absolute age scale.

To determine the absolute age of NGC\,6791, we use Monte Carlo (MC) isochrone generation to account for non-linear relationships among stellar-model input parameters. We sample broad ranges of stellar physics inputs, including composition, convection and transport, opacities, nuclear reaction rates, and external parameters. In contrast to previous absolute age-dating studies \citep{Ying2023,Ying2024,Ying2025}, we inject the observed photometric scatter directly into our synthetic CMDs (sCMD) and compare observed and synthetic CMDs one to one without oversampling. Unlike studies that use DEBs or asteroseismology as standalone age indicators, we incorporate the DEBs as explicit constraints within the same isochrone-based statistical framework. Section~\ref{sec:obs_data} describes the observational data, including the Gaia photometry, membership selection, photometric-scatter estimates, completeness correction, and adopted DEB constraints. Section~\ref{sec:iso_construct} outlines the construction of the Monte Carlo isochrone grid and details the generation of synthetic color–magnitude diagrams. Section~\ref{sec:fitting} presents the statistical fitting framework. Section~\ref{sec:results} reports the results, including the derived age and other cluster parameters, Section~\ref{sec:discussion} discusses the broader astrophysical implications, and Section~\ref{sec:conclusion} summarizes our conclusions.

\section{Observational Data} \label{sec:obs_data}

\subsection{Detached Eclipsing Binaries}\label{subsec:debs}

\citet{Brogaard2011} analyzed three detached eclipsing-binary (DEB) systems in NGC\,6791, designated V18, V20, and V80. We use DEBs here because their directly measured masses and radii provide an external distance-independent constraint on the isochrones that is complementary to CMD fitting. In their spectroscopic analysis, $T_{\rm eff}$ was measured for both components of V18 and for the primary V20.  The secondary V20 $T_{\rm eff}$ was estimated and we exclude V20s from our sample. The magnetically active system V80 is also excluded due to its comparatively uncertain radii. Thus, we adopt parameters for the primary and secondary V18 and the primary V20; The masses, radii, and luminosities adopted with uncertainties are listed in Table~\ref{tab:deb_params}.

\begin{table}[ht!]
\centering
\caption{Detached eclipsing binary (DEB) component parameters adopted from \citet{Brogaard2011}.}
\label{tab:deb_params}
\begin{tabular}{lccc}
\hline\hline
 & $M\,(M_\odot)$ & $R\,(R_\odot)$ & $L\,(L_\odot)$ \\
\hline
V18p   & $0.9955 \pm 0.0033$ & $1.1011 \pm 0.0068$ & $1.0705 \pm 0.074$ \\
V18s & $0.9293 \pm 0.0032$ & $0.9708 \pm 0.0089$ & $0.7356 \pm 0.069$ \\
V20p   & $1.0868 \pm 0.0039$ & $1.397 \pm 0.013$   & $1.7793 \pm 0.124$ \\
\hline
\end{tabular}
\end{table}

\subsection{NGC\,6791 Photometry}\label{subsec:photometry_ngc6791}

We use \emph{Gaia} DR3 photometry for NGC\,6791 \citep{GaiaCollaboration2023}. For cluster membership, we adopt the cluster center from \citet{Hunt2024} and restrict the sample to sources within three times the Jacobi radius, $3r_J$, of this center. To refine the membership selection in kinematic space, we fit a three-component Gaussian mixture model (GMM) to the proper-motion distribution $(\mu_{\alpha*},\mu_{\delta})$, identify the dominant component as the cluster population, and compute for each star its Mahalanobis distance from this component, namely the covariance-weighted distance in proper-motion space. We retain candidate members whose proper motions lie within the $2.5\sigma$ contour of the cluster Gaussian component. Our analysis focuses on main-sequence turnoff (MSTO) stars, which are highly sensitive to age but relatively insensitive to present-day mass functions \citep{Chaboyer1996}. Specifically, we select stars with $16.25 < G < 18.0$, an empirically chosen window that isolates the age-sensitive main-sequence turnoff and subgiant branch while minimizing incompleteness and contamination from less age-diagnostic regions of the CMD. Since the true \emph{Gaia} photometric uncertainties and detection completeness are complicated functions of position and local stellar density we sample photometric uncertainties directly from the NGC\,6791 data and construct completeness estimates by comparing HST and Gaia photometry, as detailed below.

\begin{figure*}[t] 
  \centering
  \includegraphics[width=\textwidth]{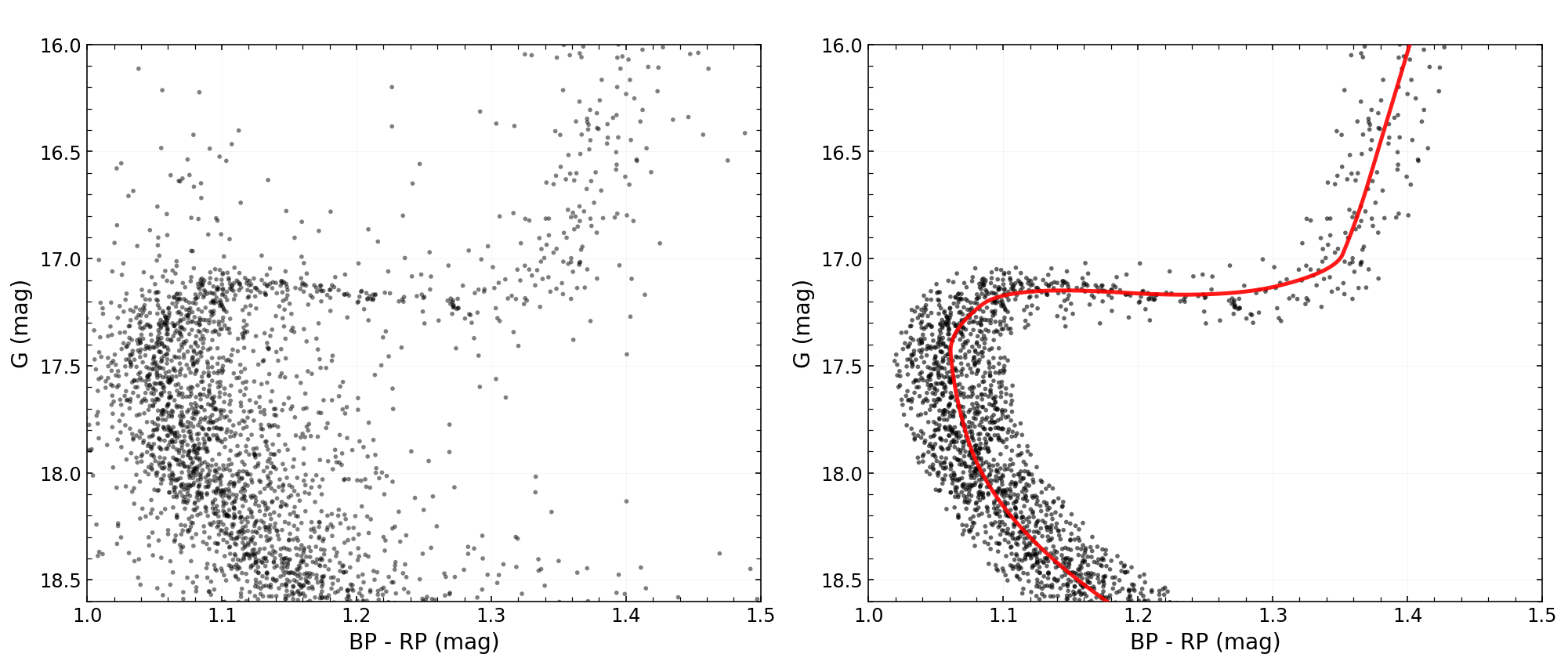}
\caption{CMDs of NGC~6791 from Gaia DR3 before (left) and after (right) ridge-line filtering with $|d_\perp|_{normalized}\leq 0.05$\,mag, the median ridge line is included in the right plot in red.}  \label{fig:ngc6791-bf-af-cuts}
\end{figure*}

Stars that closely trace the cluster’s main sequence, subgiant branch, and red giant branch are identified using Gaia $G$ magnitudes and $(BP{-}RP)$ colors. To better define these
these sequences, a ridgeline was constructed from the Color-Magnitude Diagram (CMD). The ridgeline is obtained with a dynamic-programming procedure inspired by seam carving \citep{Avidan_Shamir_2007}, in which the preferred path is guided by the local CMD density estimated from a kernel density estimate (KDE), together with penalties on large changes in slope and curvature. To accommodate the pronounced bend in the subgiant branch, we apply a temporary local function that allows the path to track this feature before relaxing back to the median sequence. The axis are normalized the axis and stars further than 0.05 away from the median ridgeline are removed. This process can be seen in Figure~\ref{fig:ngc6791-bf-af-cuts}. This process effectively removes blue stragglers and outliers that do not conform to the cluster’s princple evolutionary sequences. As a result, we obtain a cleaner, well-defined sample of stars, leading to a final dataset of 1,182 stars that accurately represents the true cluster population. The cleaned \emph{Gaia} DR3 photometric catalog is presented in Appendix~\ref{appendix:photometry}; the full electronic table spans $15.8 \leq G \leq 18.8$ mag, while the analysis below is restricted to the MSTO-focused subsample with $16.25 < G < 18.0$.

To estimate the scatter in the data (which is due to photometric errors and binaries), we bin adaptively along $G$ magnitude with minimum stars-per-bin thresholds of 50 stars, and generate merged bins to control variance at low counts. We divide the CMD into two regions: (i) $BP{-}RP \geq 1.15$ and $G < 17.5$, and (ii) $BP{-}RP < 1.15$ or $G \geq 17.5$. Each region is further subdivided into 5–10 bins, and for each bin we compute the standard deviation of the scatter perpendicular to the median ridgeline.

NGC 6791 has been observed with HST-ACS and these observations extend to considerably fainter magnitudes than the Gaia photometry.  Thus, the HST photometry can be used to estimate the incompleteness of the \emph{Gaia} photometry. Calibrated F606W and F814W photometry for NGC~6791 was retrieved from the HST UV Globular‐Cluster Survey (HUGS; GO-13297; \citealt{Nardiello2018, Soto2017}).  The footprint covers the central $\sim2\times2\,\mathrm{arcmin^{2}}$ and reaches $\mathrm{F606W}\approx25$ with quality‐flag information critical for assessing completeness. The overlap with the \emph{Gaia} data was leveraged  to derive an empirical completeness correction. We compared HST and Gaia photometry by associating F606W with Gaia $G_{\mathrm{BP}}$ and F814W with $G_{\mathrm{RP}}$, while Gaia $G$ is approximately equivalent to a combination of the two \citep{Bedin2005,Evans2018}. Quality indicators in the HUGS dataset (quality-fit, sharpness, and saturation flags) show that the HST photometry remains reliable to $m \sim 22$–23 in both F606W and F814W far beyond what we use in our Gaia data. By analyzing the overlapping sky region, we compared the number of stars detected in \emph{Gaia} and HST as a function of magnitude and used their ratio as an empirical estimate of \emph{Gaia}'s completeness relative to HST. This correction is primarily relevant for the fainter portion of the sample, where missed \emph{Gaia} detections become non-negligible. In the brighter subgiant- and red-giant-branch bins, both surveys are expected to be effectively complete, and the raw \emph{Gaia}/HST count ratios remain close to unity but become visibly noisy because each bin contains only of order 10-15 stars. For that reason, we do not interpret the bright-bin fluctuations as evidence for real incompleteness and instead set the completeness to unity in these phases. Cross-matching the Gaia and HST catalogs by position and approximate magnitude enables a transformation between the Gaia and HST filter systems, providing a robust completeness function that can then be applied in our analysis. The final completeness is listed in Table~\ref{tab:gaia_completeness_simplified}.

\begin{table}[ht]
\centering
\begin{tabular}{cc}
\hline
\textbf{G-band Bin Center} & \textbf{Completeness} \\
\hline
15.25 & 1.000 \\
15.75 & 1.000 \\
16.25 & 1.000 \\
16.75 & 1.000 \\
17.25 & 0.930 \\
17.75 & 0.838 \\
18.25 & 0.656 \\
18.75 & 0.476 \\
19.25 & 0.127 \\
\hline
\end{tabular}
\caption{Gaia G-band completeness as a function of magnitude bin center. In the $2\times2$~arcmin overlap region we compare star counts in 0.5‑mag bins between HUGS (assumed complete to F606W$<22$) and \emph{Gaia}.  The resulting completeness fraction $C(G)$ is capped at unity and forced to 1.0 for the red‑giant branch (RGB).
}
\label{tab:gaia_completeness_simplified}
\end{table}

\section{Isochrone Construction}\label{sec:iso_construct}

We need an isochrone library that is fast and flexible enough to marginalize over key stellar-physics uncertainties. Pre-computed isochrone grids generally do not span the range of input physics we require, so we generate models with the Dartmouth Stellar Evolution Program (DSEP) \citep{Dotter2008}, adopting literature-based priors summarized in Table~\ref{tab:MCparams}. The composition priors were anchored to published measurements for NGC\,6791. For $[\mathrm{Fe/H}]$ and $[\alpha/\mathrm{Fe}]$, we adopted ranges based on high-resolution spectroscopic studies of cluster members, with the $[\alpha/\mathrm{Fe}]$ prior chosen to reflect the approximately solar to mildly enhanced $\alpha$-element abundances reported across the literature, while for the initial helium mass fraction we adopted a deliberately broad uniform prior, $Y_{0}=0.27$--$0.33$, centered on previous DEB- and asteroseismic-based estimates near $Y\approx0.30$ but widened to allow for residual systematic uncertainty \citep{Villanova2018,Linden2017,Cunha2015,Donor2018,Brogaard2012,McKeever2019}. DSEP uses the one-dimensional mixing-length formalism of \citet{1965ApJ...142..841H}, the standard approach in most stellar-structure and evolution codes. Because a solar-calibrated mixing length is not adequate for all stars \citep{Joyce2018,Guenther2000}, a broad prior is adopted for the mixing-length parameter $\alpha_{\mathrm{mlt}}$  (Table~\ref{tab:MCparams}). Another source of uncertainty is convective overshoot at both the base of the convective envelope and the boundary of the convective core.  Literature-motivated ranges are adopted for the corresponding overshoot parameters \citep{Claret2004,Demarque2004,Pietrinferni2004,Mowlavi2012}, and note that independent asteroseismic analyses also motivate nonzero overshoot and constrain envelope and core boundary mixing \citep{Lindsay2022,Lindsay2024}. We do not discuss the remaining physics parameters in detail here, but they can be found in \citet{Dotter2008} and \citet{Ying2023}.

\begin{deluxetable*}{llll}
\tablecaption{Adopted priors for Monte Carlo isochrone inputs, distance modulus, and reddening.\label{tab:MCparams}}
\tablehead{
\colhead{Parameter} & \colhead{Prior} & \colhead{Adopted range} & \colhead{Reference(s)}
}
\startdata
\multicolumn{4}{c}{\it Composition} \\
\hline
$[\mathrm{Fe/H}]$ & Normal & $+0.31 \pm 0.11$ &
\citep{Villanova2018,Linden2017,Cunha2015,Donor2018} \\
$[\alpha/\mathrm{Fe}]$ & Normal & $0.06 \pm 0.08$ &
\citep{Villanova2018,Linden2017,Donor2018} \\
$Y_{0}$ (Initial He) & Uniform & $0.27$--$0.33$ &
\citep{McKeever2019,Brogaard2012} \\
\hline
\multicolumn{4}{c}{\it Convection and transport} \\
\hline
$\alpha_{\rm MLT}$ (mixing length) & Uniform & $1.0$--$2.5$ & {N/A} \\
Diffusion scale (metals) & Uniform & $0.5$--$1.3$ & \citep{Thoul1994} \\
Diffusion scale (He) & Uniform & $0.5$--$1.3$ & \citep{Thoul1994} \\
Envelope overshoot & Uniform & $0.0$--$0.2$ & {N/A} \\
Core overshoot & Uniform & $0.0$--$0.2$ & {N/A} \\
Surface boundary condition & Trinary & $1/3$ each &
\citep{Eddington1926,KrishnaSwamy1966,Hauschildt1999} \\
\hline
\multicolumn{4}{c}{\it Microphysics} \\
\hline
Low-$T$ opacity scale & Uniform & $0.7$--$1.3$ & \citep{Ferguson2005} \\
High-$T$ opacity scale & Normal & $1.00 \pm 0.03$ & \citep{Iglesias1996} \\
Plasma neutrino-loss scale & Normal & $1.00 \pm 0.05$ & \citep{Haft1994} \\
Conductive opacity scale & Normal & $1.00 \pm 0.20$ & \citep{Hubbard1969,Canuto1970} \\
\hline
\multicolumn{4}{c}{\it Nuclear reaction rates (S-factors; keV\,b)} \\
\hline
$p(p,e^+\nu)\,^2$H & Normal & $(4.07 \pm 0.04)\times10^{-22}$ &
\citep{Acharya2016,Marcucci2013} \\
$^3$He($^3$He,$2p)^4$He & Normal & $5150 \pm 500$ & \citep{Adelberger2011} \\
$^3$He($^4$He,$\gamma)^7$Be & Normal & $0.54 \pm 0.03$ & \citep{deBoer2014} \\
$^{12}$C($p,\gamma)^{13}$N & Normal & $1.45 \pm 0.50$ & \citep{Xu2013} \\
$^{13}$C($p,\gamma)^{14}$N & Normal & $5.50 \pm 1.20$ & \citep{Chakraborty2015} \\
$^{14}$N($p,\gamma)^{15}$O & Normal & $3.32 \pm 0.11$ & \citep{Marta2011} \\
$^{16}$O($p,\gamma)^{17}$F & Normal & $9.40 \pm 0.80$ & \citep{Adelberger2011} \\
\hline
\multicolumn{4}{c}{\it External parameters} \\
\hline
$(m{-}M)$ (apparent DM) & Uniform & $13.23$--$13.50$ &
\citep{Brogaard2011,VandenBerg_2014,Chaboyer1999} \\
$E(B{-}V)$ & Uniform & $0.10$--$0.20$ &
\citep{Brogaard2011,VandenBerg_2014,Chaboyer1999,Linden2017} \\
\enddata
\tablecomments{``Scale'' parameters multiply the corresponding default input physics in the stellar models. The surface boundary condition is sampled with equal probability among the three listed prescriptions.}
\end{deluxetable*}

\begin{figure}[!ht] 
  \centering
  \includegraphics[width=8cm]{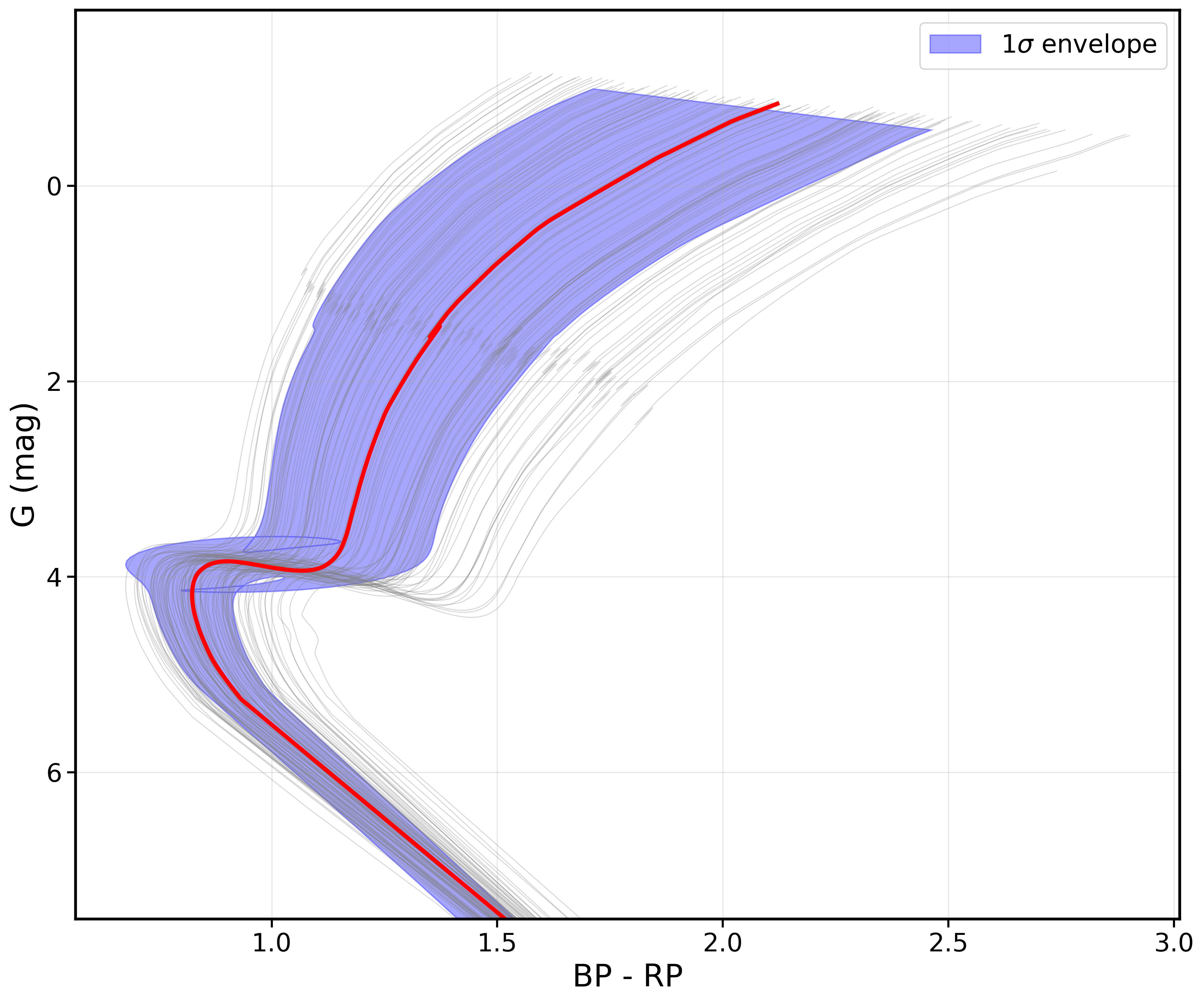}
  \caption{Ensemble of 10{,}000 Monte Carlo 8.5 Gyr isochrones generated for this work (gray). The red curve shows the medoid isochrone, defined as the member of the ensemble that minimizes the total cumulative distance to all other isochrones in the $(BP{-}RP,\,G)$ plane. The shaded blue region marks the 68\% confidence interval.}
  \label{fig:mc_isochrone_ensemble_8.5Gyr}
\end{figure}

A library of \(10{,}000\) Monte Carlo isochrone sets designed to cover the evolutionary phases spanned by our CMD fitting window was constructed. For each set, stellar-evolution parameters are drawn from the priors in Table~\ref{tab:MCparams} using Sobol sampling, and isochrones are then built using the Equal Evolutionary Point (EEP) formalism \citep{Dotter2016}. Our track grid spans initial masses from \(0.1\) to \(3\,M_{\odot}\). For the low-mass models \texttt{FreeEOS-2.2.1} is adopted \citep{Irwin2012}, while the higher-mass models use an equation of state including the Debye--H\"uckel correction \citep{Chaboyer_Kim_1995}. We compute isochrones over ages \(4\)–\(12\)~Gyr in \(0.1\)~Gyr increments, with 400 EEPs per isochrone. Each Monte Carlo set therefore contains 41 isochrones (one per age) for testing. To place the models in the \emph{Gaia} observational plane, bolometric corrections from the MIST tables are adopted to convert the theoretical stellar parameters to $G$, $G_{\rm BP}$, and $G_{\rm RP}$ magnitudes \citep{Dotter2016,Choi2016}. Figure~\ref{fig:mc_isochrone_ensemble_8.5Gyr} shows the Monte Carlo ensemble of the 8.5~Gyr isochrones used in this work, highlighting how marginalized systematic uncertainties in the stellar evolution models broaden the predicted isochrone sequence in CMD space.

\subsection{Synthetic Color Magnitude Diagrams}\label{subsec:scmd}
Each theoretical isochrone in age, apparent distance modulus, and reddening is used to generate a synthetic color magnitude diagram (sCMD) of NGC 6791 for comparison with its Gaia CMD. For each isochrone, the sCMD is created by drawing 1182 stars, the total star count in our final CMD sample \ref{sec:obs_data} using the following procedure:

\begin{enumerate}
\item A present-day stellar mass is randomly selected from a single-slope power-law mass function with slope $-1.02$. Because the simulated CMD spans only a narrow mass range around the main-sequence turnoff and subgiant branch, the exact choice of slope has little impact on the results (see discussion below), and a value of $-1.02$ is adopted for convenience. The mass is mapped to Gaia $G$, $BP$, and $RP$ magnitudes via MIST bolometric corrections accessed through \texttt{fidanka} \citep{Dotter2016, Choi2016, Boudreaux_fidanka_2023}, then shifted to apparent values using Gaia DR3 extinction coefficients and the specified reddening; extinction is applied band-by-band and the distance modulus is treated consistently so that extinction is not double-counted.
\item Completeness-corrected target star counts are assigned in fixed $G$-magnitude bins using the empirical completeness function in Table~\ref{tab:gaia_completeness_simplified}. For each bin, we infer the underlying number of stars from the observed count and the completeness fraction, sample that number from the isochrone, and then apply incompleteness by randomly removing stars according to the same completeness fraction.
\item Empirical photometric scatter is injected perpendicular to the local isochrone in the $G$ versus $(BP{-}RP)$ normalized plane. The local isochrone slope is estimated, and stars receive $G$-dependent, perturbations perpendicular to the isochrone. To assign the perturbations, we divide the isochrone into two regions: (i) $BP{-}RP \geq 1.15$ and $G < 17.5$, and (ii) $BP{-}RP < 1.15$ or $G \geq 17.5$, with separate bins defined in each region for sampling the injected errors.
\item The perturbed Gaia magnitudes ($G_{\rm obs}$, $BP_{\rm obs}$, $RP_{\rm obs}$) and color $(BP{-}RP)_{\rm obs}$ define the final sCMD used for quantitative comparison to the observed CMD.
\end{enumerate}

The adopted distance modulus $\mu$ and reddening $E(B\!-\!V)$ are listed in Table~\ref{tab:MCparams}. After applying these values, each theoretical isochrone yields 1182 simulated stars. While \citet{Ying2023,Ying2024,Ying2025} typically generated large oversampled catalogs (10–20$\times$ the observed star count), we simulate only a 1$\times$ sample so that changes in the test statistic reflect model–data mismatch rather than sample-size scaling.  Finally, we apply completeness only as a function of magnitude, consistent with our uncertainty model.

A $10,000$ realization robustness test varying the assumed power law mass function slope from $\alpha = -1.02$  to $\alpha = 0.0$ (estimated from the HST CMD; \citealt{Dalessandro2015, King2005})  shifts the median 2D KS statistic by only $\approx 1 - 2\times 10^{-3}$, corresponding to a few percent of the bootstrap calibrated KS scale ($\mathrm{median}\approx 2.6\times 10^{-2}$), and we therefore treat the CMD weights as insensitive to plausible mass function slopes in the MSTO fitting window. Including an unresolved binary fraction $f_{\rm b} = 0.25$ \citep{Bedin2008WD} in a $10,000$ realization test shifts the median 2D KS statistic by $\approx 1 \times 10^{-3}$ (again only a few percent of the bootstrap calibrated KS scale), so we do not model binaries explicitly and regard them as a secondary systematic for the CMD region analyzed here.

\section{Isochrone Fitting}\label{sec:fitting}

In general, we generate $\approx 2.26\times 10^{8}$ synthetic CMD realizations that span our Monte Carlo isochrone sets, distance modulus, and reddening, and we require a quantitative statistic to evaluate how well each sCMD matches the observed \emph{Gaia} CMD. Recent absolute age studies have compared full-CMD fitting approaches based on adaptive binning and multivariate distribution tests \citep[e.g.,][]{Ying2023,Ying2024,Ying2025}. In our application, the two-dimensional Kolmogorov–Smirnov (2D KS) framework is preferred because it is less sensitive to choices of binning and is comparatively robust to photometric uncertainty modeling, which is important here because we work with \emph{Gaia} photometry rather than HST artificial-star tests and we inject scatter empirically using ridge-line based estimates. We therefore adopt the 2D KS test throughout this analysis.

The 2D KS test generalizes the one-dimensional Kolmogorov-Smirnov statistic to two variables by comparing the empirical cumulative distribution functions (ECDFs) of two samples. Following the original formulation of the 2D KS test \citet{Peacock1983} and the commonly used refinement of \citet{Fasano1987}, we consider $n$ observed stars and $m$ synthetic stars, with ECDFs $F_n(x,y)$ and $G_m(x,y)$, respectively, where each ECDF gives the fraction of stars with coordinates less than or equal to $(x,y)$. The statistic is defined as
\begin{equation}
D_{n,m} = \sup_{x,y} \left| F_n(x,y) - G_m(x,y) \right| .
\label{eq:2dks}
\end{equation}
Operationally, the CMD morphology is evaluated by comparing the observed and synthetic samples through quadrant counts around trial points in the CMD. For each trial location $(x,y)$, we compute the fractions of stars lying in each of the four quadrants relative to that point for both samples, evaluate the absolute differences between the corresponding quadrant fractions, and retain the largest such difference. Efficient evaluation over large samples can be carried out with divide-and-conquer ranking methods popularized by \citet{Bentley1980}. In this work, we compute this maximum difference over the CMD region used for fitting and adopt it as our CMD goodness-of-fit metric.

\begin{figure*}[!ht] 
  \centering
  \includegraphics[width=\textwidth]{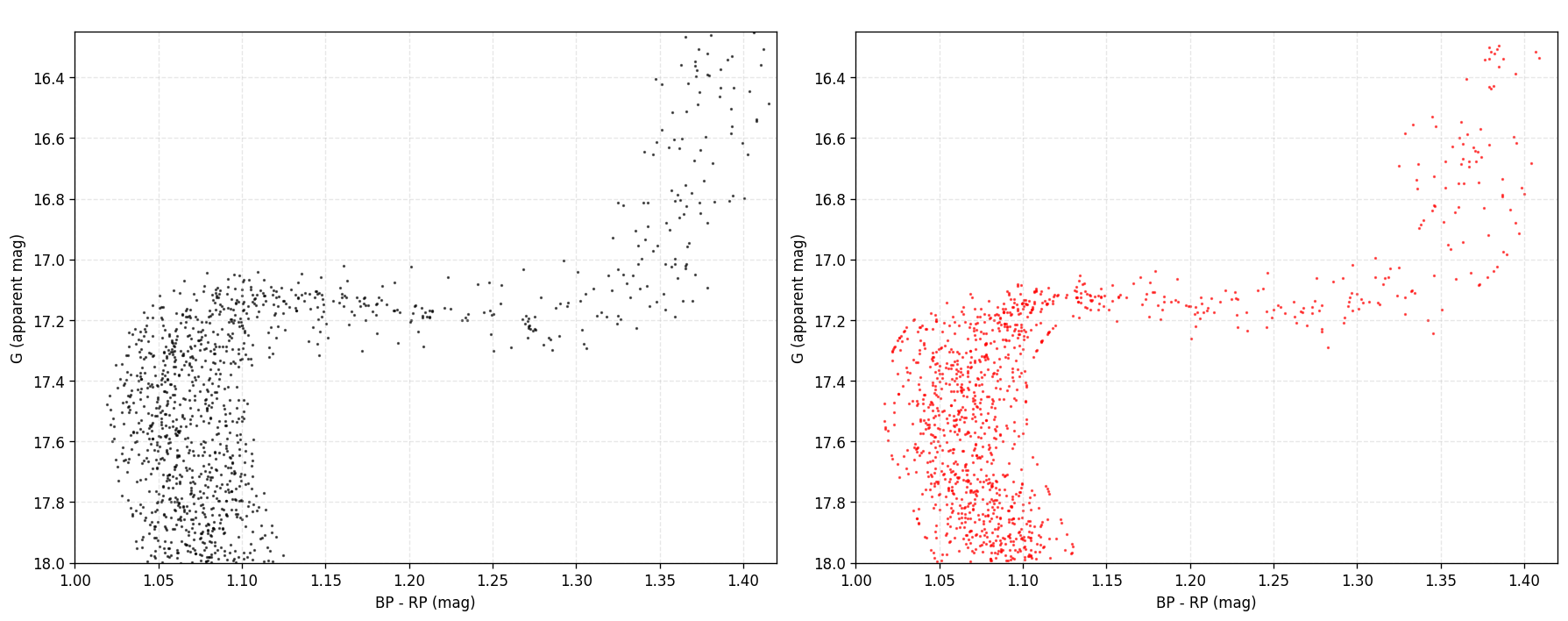}
    \caption{Comparison between the observed color--magnitude diagram (CMD, left panel) and a well fitting simulated CMD (right panel) obtained from the 2D Kolmogorov--Smirnov (KS) test. The observed CMD is shown in black, while the simulated CMD is displayed in red. The parameters are: age = 8.4 Gyr, $\mu = 13.24$, and $E(B-V) = 0.19$.}
    \label{fig:cmd_obs_vs_sim}
\end{figure*}

\begin{figure}[!ht] 
  \centering
  \includegraphics[width=8cm]{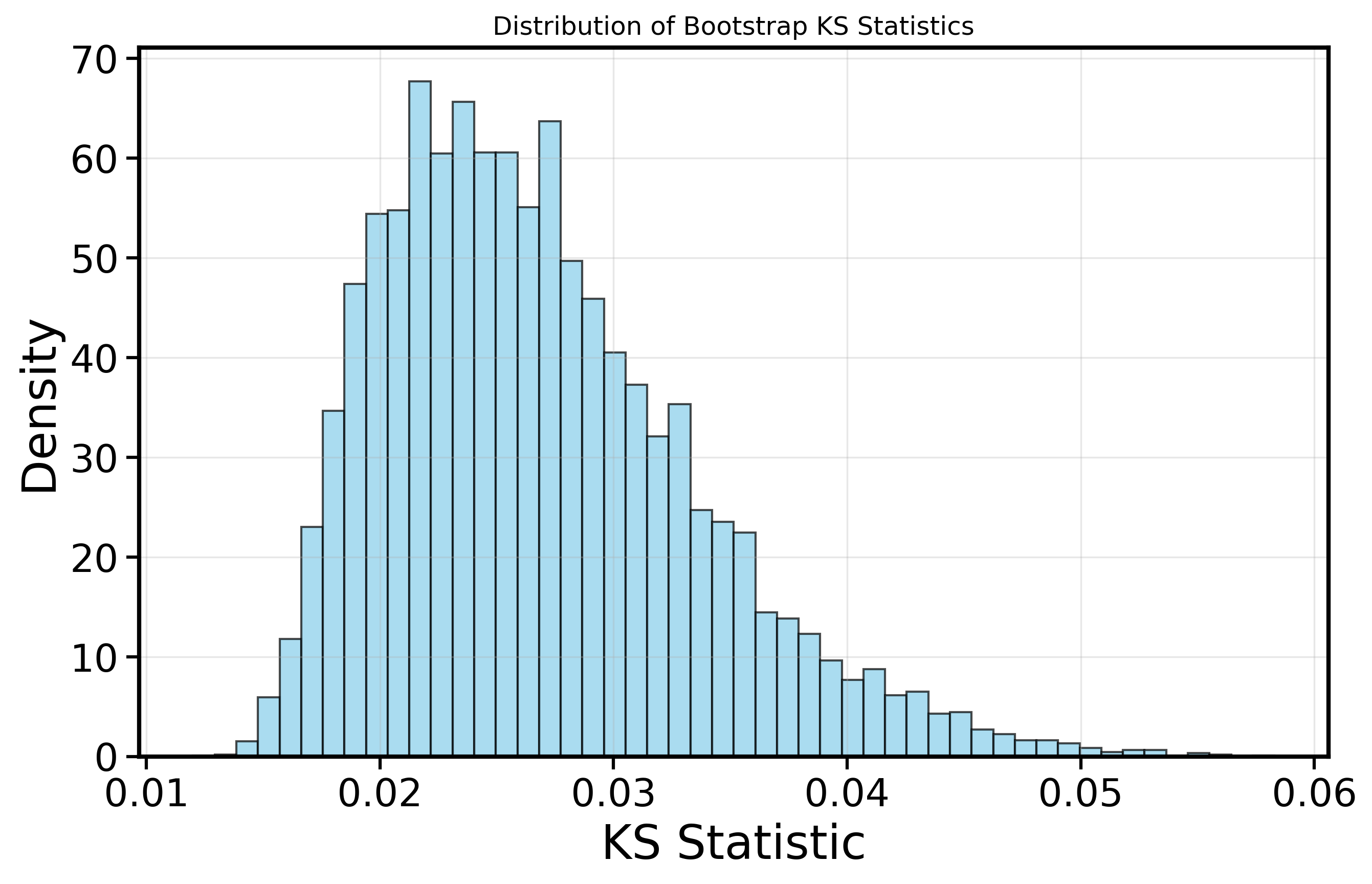}
  \caption{Bootstrap resampling for the 2D-KS method. The empirical 2D-KS distribution from 10,000 bootstrap resampling.}
  \label{fig:ks_statistic_distribution_scmd}
\end{figure}

Because the sampling distribution of the 2D KS statistic depends on the underlying parent distribution, we cannot map $D_{n,m}$ to a $p$-value using a generic analytic mapping. Instead, we calibrate the statistic with a resampling bootstrap tailored to our sCMD generation. We take our representative best-fitting sCMD realization at $(t,\mu,E(B{-}V)) = (8.4,\mathrm{Gyr},13.24,0.19)$ (Figure~\ref{fig:cmd_obs_vs_sim}) and treat it as the reference parent distribution under the null hypothesis. We then draw 10{,}000 resampled sCMDs from this same parent model, rerun the 2D KS calculation for each resample, and thereby obtain an empirical reference distribution that captures the intrinsic stochasticity introduced by IMF sampling, completeness application, and the injected photometric scatter. Each model sCMD is assigned a probability weight by mapping its measured $D_{n,m}$ onto this bootstrap-calibrated reference distribution. The resulting bootstrap-calibrated distribution of 2D KS statistics is shown in Figure~\ref{fig:ks_statistic_distribution_scmd}.

Given the full grid size, we apply an initial screening step to reduce the number of sCMDs carried into the bootstrap-calibrated 2D KS stage. For each Monte Carlo isochrone set, we first rank the candidate sCMDs using a kernel-density based comparison between the observed CMD and each simulated CMD on a common $(BP{-}RP,G)$ grid. Specifically, we compute two-dimensional kernel density estimates (KDEs) for the observed and simulated CMDs, quantify their separation using the Jensen–Shannon distance, a symmetric measure of distributional difference \citep{LinJ1991}, and retain only the best-matching sCMDs per age for subsequent analysis. This filter is used as a computational throttle and is not intended to replace the 2D KS statistic, which remains our primary goodness-of-fit metric for CMD morphology.

\begin{figure}[!ht]
  \centering
  \includegraphics[width=\textwidth]{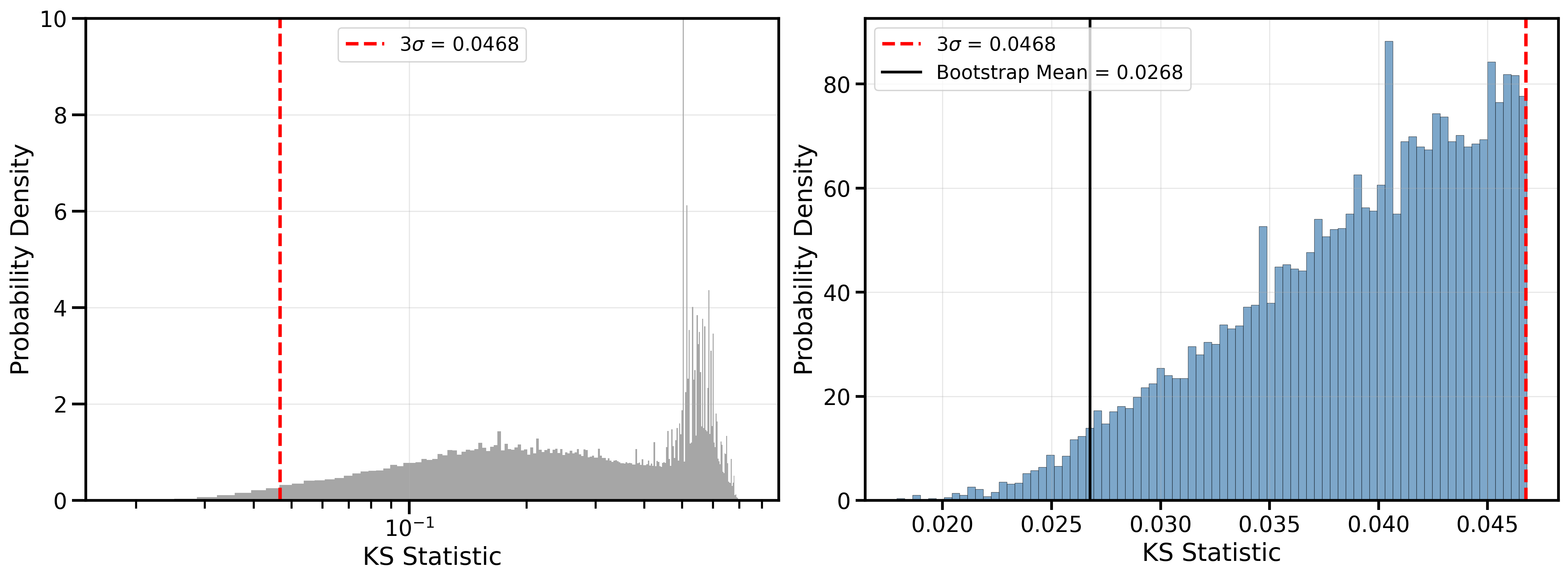}
  \caption{Distribution of 2D KS statistics for the synthetic CMD realizations. The left panel shows the full distribution for all $\sim 2.26\times10^{8}$ realizations, demonstrating that the vast majority of models provide poor fits to the observed CMD. The right panel shows a zoomed view of the low-KS tail, with the red dashed line marking the adopted $3\sigma$ selection threshold derived from the bootstrap distribution and the black vertical line marking the bootstrap mean. Realizations to the left of the red dashed line are retained for the final weighted analysis.}
  \label{fig:ks_statistic_3sigma_zoom}
\end{figure}

Figure~\ref{fig:cmd_obs_vs_sim} shows an example of an sCMD that provides a close match to the observed CMD in the fitting window. Figure~\ref{fig:ks_statistic_3sigma_zoom} shows the distribution of 2D KS statistics for the synthetic CMD realizations. The full distribution demonstrates that most models are poor fits to the data, while the zoomed low-KS tail highlights the subset within the adopted $3\sigma$ threshold of the bootstrap mean that is retained for the final weighted analysis. After this KS-based selection, and after keeping only the best distance-modulus and reddening combination for each isochrone, we retain $13{,}932$ well-fitting sCMDs for the remainder of the analysis.

To test the performance of the sCMD fitting procedure, we carried out a controlled mock-recovery test using an isochrone drawn from an independent stellar-evolution library. Specifically, we generated a mock \emph{Gaia}-like CMD from a MIST isochrone \citep{Dotter2016,Choi2016} with input parameters $t_{\rm input}=8.3$~Gyr, $[\mathrm{Fe/H}]_{\rm input}=+0.30$, $[\alpha/\mathrm{Fe}]_{\rm input}=0.0$, $Y_{\rm input}=0.2930$, $(m-M)_{\rm input}=13.35$, and $E(B-V)_{\rm input}=0.20$. The mock CMD was constructed using the same number of stars, mass-function sampling, completeness correction, and empirical photometric-scatter prescription adopted from our analysis. We then analyzed this mock CMD with the same DSEP-based sCMD recovery pipeline used for the observed cluster.

The recovered sCMD-only posterior has median parameter values of $t=8.48\pm0.69$~Gyr, $[\mathrm{Fe/H}]=+0.280\pm0.092$, $Y=0.300\pm0.017$, $[\alpha/\mathrm{Fe}]=0.0438\pm0.0953$, $(m-M)=13.256\pm0.058$, and $E(B-V)=0.177\pm0.027$, where the uncertainties denote the standard deviations of the recovered distributions. The recovered age differs from the input MIST age by only $+0.18$~Gyr, which is small compared with both the mock-recovery uncertainty and the final age uncertainty for NGC\,6791. This test shows that the sCMD pipeline does not introduce a significant age bias when applied to a mock CMD generated from an independent stellar-evolution library. We also repeated the mock-recovery experiment using a DSEP input isochrone with $t_{\rm input}=8.3$~Gyr, $(m-M)_{\rm input}=13.280$, and $E(B-V)_{\rm input}=0.200$; the recovered sCMD-only posterior gives $t=8.49\pm0.62$~Gyr, $(m-M)=13.251\pm0.050$, and $E(B-V)=0.181\pm0.020$.

The recovered distance modulus and reddening are offset from the input values by $-0.094$~mag and $-0.023$~mag, respectively. These offsets likely reflect the combined effects of finite sCMD sampling, injected photometric scatter, completeness corrections, and cross-library differences in the CMD morphology. They also illustrate the covariance among age, distance modulus, and reddening in CMD-based fitting. We therefore interpret this test as evidence that the age recovery is robust at the level required for our analysis, while retaining caution that external parameters such as distance modulus and reddening may be more sensitive to photometric systematics and library dependent stellar physics systematics.

\subsection{DEB isochrone fitting and bootstraps}\label{subsec:cal_deb_fits}

In addition to the CMD-based statistic, we include a complementary, largely distance- and reddening-independent check on our Monte Carlo isochrone sets using detached eclipsing binaries (DEBs) in NGC\,6791. These external constraints help identify Monte Carlo realizations that reproduce key stellar properties, reducing our reliance on the CMD fit alone. We adopt a nearest-point goodness-of-fit approach similar to that used in related isochrone benchmarking work \citep[e.g.,][]{OMalley2017,Ying2024,Ying2025}.

We adopt the DEB component parameters from \citet{Brogaard2011} listed in Table~\ref{tab:deb_params}. In the present analysis we use the primary and secondary components of V18 and the primary component of V20. Although masses, radii, and luminosities are available for these systems, we found that the $M$--$L$ statistic provides the most stable constraint across the full age range explored here. Radius-based combinations become much more sensitive near the main-sequence turnoff and subgiant regime, where small local differences in isochrone morphology can lead to disproportionately large changes in the DEB $\chi^2$ statistic.

For each DEB component $j$ and isochrone of age $t$, we evaluate the isochrone luminosity at the observed DEB mass $M_j$ and define the per-star mismatch as
\begin{equation}
\chi^2_{\mathrm{DEB},j}(t) =
\left(\frac{\log L_j - \log L_{\mathrm{iso}}(M_j,t)}{\sigma_{\log L,j}}\right)^2,
\label{eq:deb_chi2_point}
\end{equation}
where $\sigma_{\log L,j} = \sigma_{L,j}/(L_j \ln 10)$. The total DEB statistic for a given isochrone is then
\begin{equation}
\chi^2_{\mathrm{DEB}}(t) =
\sum_{j=1}^{N_{\mathrm{DEB}}} \chi^2_{\mathrm{DEB},j}(t).
\label{eq:deb_chi2_total}
\end{equation}
This construction yields a single scalar DEB goodness-of-fit statistic for each isochrone while preserving the direct mass anchoring provided by the DEB observations.

\begin{figure}[!ht]
  \centering
  \includegraphics[width=12cm]{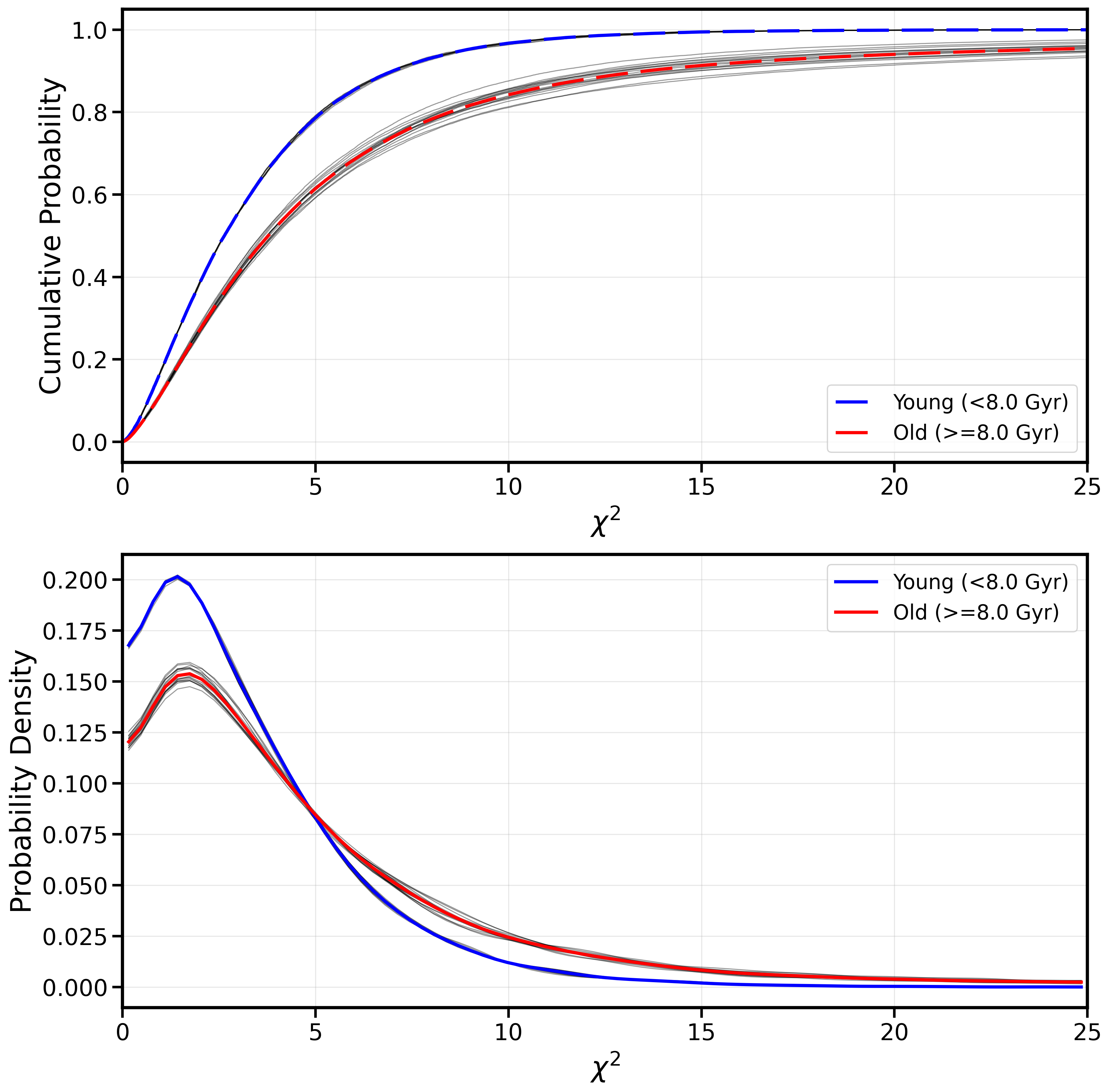}
  \caption{Bootstrap calibration of the DEB nearest-point statistic, $\chi^2_{\mathrm{DEB}}$, for the adopted $M$--$L$ fitting scheme using the V18p, V18s, and V20p components. Thin gray curves show the empirical distributions obtained from approximately 100 individual well-fitting reference isochrones, while the colored curves show the combined calibrations used in the analysis. The top panel gives the cumulative distribution functions and the bottom panel the corresponding probability density functions. The bootstrap realizations separate into two age-dependent families, one for isochrones younger than 8.0 Gyr (blue) and one for isochrones with ages $\geq 8.0$ Gyr (red); these two empirical calibrations are used to map each model's $\chi^2_{\mathrm{DEB}}$ value onto a DEB weight.}
  \label{fig:calibration_weights_deb}
\end{figure}

The sampling distributions of our nearest-point statistics are not guaranteed to follow an ideal $\chi^2$ law because they depend on the local isochrone geometry and on how measurement errors project into the chosen observable space. We therefore calibrate our DEBs weights using empirical reference distributions generated by Monte Carlo resampling. 

We select a well-fitting reference isochrone and generate $10{,}000$ mock DEB realizations by drawing synthetic stars near the observed DEB masses and perturbing their luminosities with Gaussian noise using the uncertainties in Table~\ref{tab:deb_params}. For each realization we recompute the coeval DEB statistic $\chi^2_{\mathrm{DEB}}$ (Equation~\ref{eq:deb_chi2_total}), producing an empirical null distribution for $\chi^2_{\mathrm{DEB}}$. To make this calibration robust against modest changes in the reference model, we repeated this procedure for approximately 100 well fitting isochrones spanning the relevant age and parameter space. 

The resulting bootstrap distributions separate naturally into two age regimes, one for isochrones younger than 8 Gyr and one for isochrones with ages $\geq 8$ Gyr. This split is driven primarily by V20p, whose position near the main-sequence turnoff causes the DEB statistic to become highly sensitive to small changes in isochrone morphology. We therefore construct separate empirical cumulative distribution functions for these two regimes and use them to map each tested isochrone's $\chi^2_{\mathrm{DEB}}$ value onto a DEB weight, $w_{\mathrm{DEB}}$. The resulting bootstrap calibration is illustrated in Figure~\ref{fig:calibration_weights_deb}.

Finally, for each model tested, the measured $\chi^2_{\mathrm{DEB}}$ value is mapped onto the corresponding bootstrap reference distribution to obtain a probability weight $w_{\mathrm{DEB}}$. This weight is then multiplied with the CMD-based weight to define the overall isochrone weight,
\begin{equation}
w_{\mathrm{iso}} = w_{\mathrm{sCMD}} \, w_{\mathrm{DEB}} .
\label{eq:iso_weight}
\end{equation}
In this way, the CMD provides the primary population-level constraint, while the DEBs supply an external, distance- and reddening-independent consistency check. The resulting $w_{\mathrm{iso}}$ values are then used to infer the age and other cluster parameters of NGC\,6791.

All synthetic CMD generation, ridgeline filtering, bootstrap calibration, and fitting steps were implemented in the \texttt{isochronetoolbox} package \citep{isochronetoolbox_github_2026}.

\section{Results}\label{sec:results}

\subsection{Age Estimates}\label{subsec:age_est}

Using the combined isochrone weights $w_{\mathrm{iso}}$ (Equation~\ref{eq:iso_weight}), we construct a marginalized, weighted age distribution from the full Monte Carlo grid. We first retain only those sCMD realizations whose 2D KS statistics fall within the resampled bootstraps $3\sigma$ acceptance region, ensuring that the retained models are statistically consistent with the reference distribution. We then renormalize the remaining weights so that $\sum_i w_{\mathrm{iso},i}=1$ and marginalize over all the distance, redennening and MC physics inputs to obtain the one-dimensional age distribution $p(t)$. 

\begin{figure}[!ht]
  \centering
  \includegraphics[width=15cm]{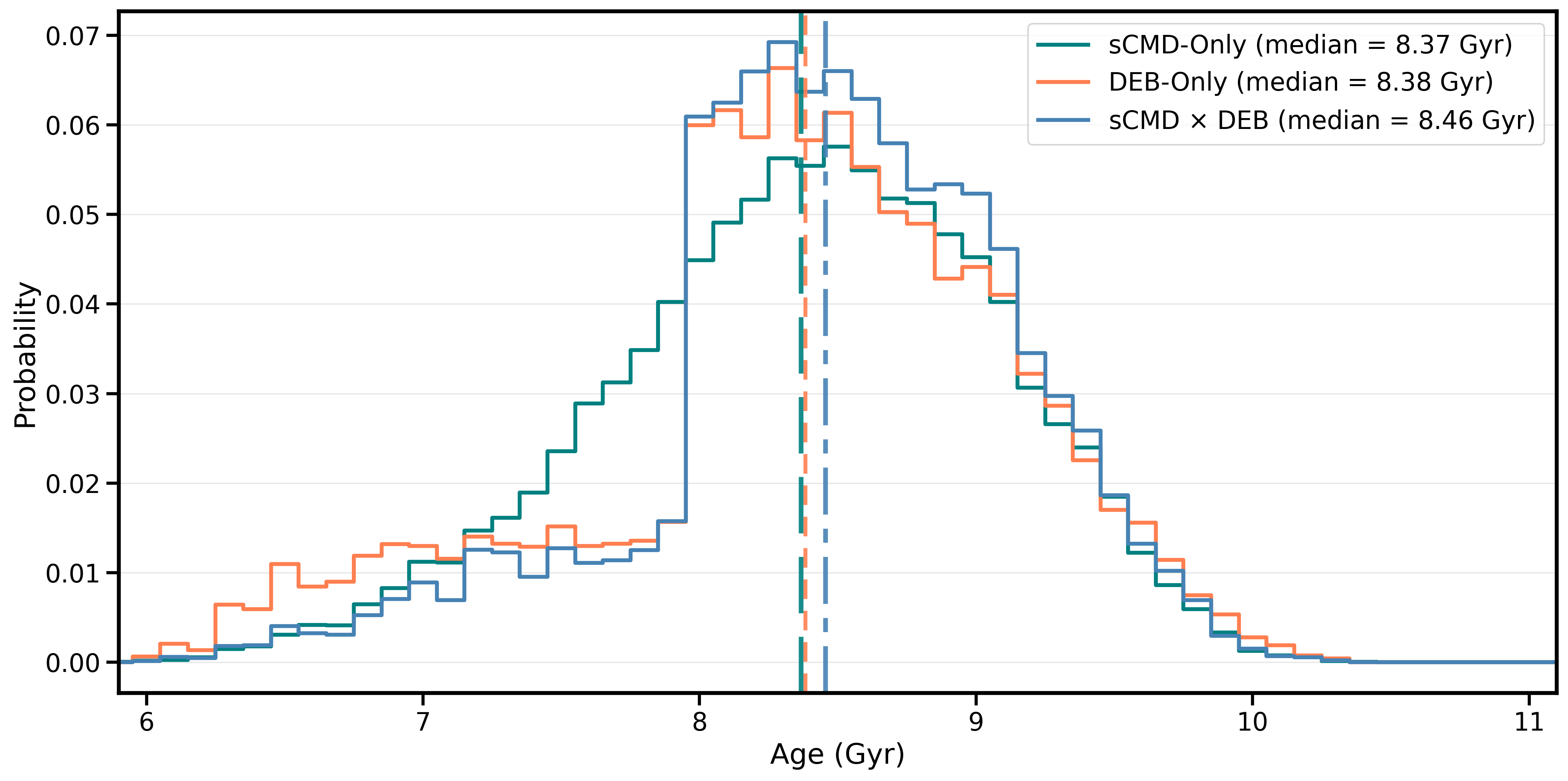}
  \caption{Marginalized posterior age distributions for NGC 6791 from the sCMD-only, DEB-only, and joint sCMD$\times$DEB analyses. Vertical dashed lines mark the posterior medians. The joint sCMD$\times$DEB solution, which is adopted throughout the remainder of this work, yields a median age of $8.46 \pm 0.66$ Gyr.}
  \label{fig:weighted_age_distribution}
\end{figure}

The resulting age constraints are shown in Figure~\ref{fig:weighted_age_distribution}, which compares the sCMD-only, DEB-only, and joint sCMD$\times$DEB posteriors. The age posteriors in Figure~\ref{fig:weighted_age_distribution} differ in shape: the sCMD-only posterior is approximately symmetric, while the DEB-only and sCMD $\times$ DEB posteriors are more asymmetric because the DEB constraints are sensitive near the MSTO and carry stronger age leverage. We therefore adopt the posterior median and the 16th–84th percentile interval as our primary summary statistics. The sCMD-only posterior gives an age of $8.37\pm0.70$ Gyr, while the DEB-only posterior gives $8.38\pm0.78$ Gyr. Our baseline result comes from the joint sCMD$\times$DEB analysis, which yields an age of $8.46\pm0.66$ Gyr.

The extended tails in Figure~\ref{fig:weighted_age_distribution} primarily reflect residual degeneracies between age, distance modulus, and reddening, along with the broadening of isochrone morphology produced by the Monte Carlo physics ensemble. Previous literature results and our summary constraints for age, $[\mathrm{Fe/H}]$, $Y$, $(m-M)_V$, and $E(B{-}V)$ are listed in Table~\ref{tab:ngc6791_results}. Our inferred age is consistent with previous determinations within the quoted 68\% interval, while providing a more comprehensive absolute-age constraint through the joint use of CMD and DEB information within a fully marginalized Monte Carlo framework. Differences in $\mu$ and $E(B{-}V)$ likely reflect updated membership selection and the use of our specific photometric dataset.

\begin{deluxetable*}{lccccc}[!ht]
\tablecaption{Summary of Results \label{tab:ngc6791_results}}
\tablehead{
\colhead{Source} &
\colhead{Age (Gyr)} &
\colhead{[Fe/H]} &
\colhead{$Y$} &
\colhead{$(m\!-\!M)_V$} &
\colhead{$E(B-V)$}
}
\startdata
\citet{Brogaard2012} &
$8.3 \pm 0.3$ &
$+0.29 \pm 0.09$\tablenotemark{a} &
$0.30 \pm 0.01$ &
$13.51 \pm 0.06$ &
$0.14 \pm 0.02$ \\
\citet{Chaboyer1999} &
$8.0 \pm 0.5$ &
$+0.40 \pm 0.10$ &
\dots\tablenotemark{b} &
$13.42$ &
$0.10$ \\
\citet{VandenBerg_2014} &
$8.5$ &
$+0.30$ &
$0.28$ &
$13.05$\tablenotemark{c} &
$0.16$ \\
\citet{McKeever2019} &
$8.2 \pm 0.3$ &
\dots\tablenotemark{d} &
$0.297 \pm 0.003$ &
\dots\tablenotemark{d} &
\dots\tablenotemark{d} \\
\hline
\multicolumn{6}{c}{\textbf{This work}} \\
\hline
sCMD-only &
$8.37 \pm 0.70$ &
$+0.277 \pm 0.094$ &
$0.2991 \pm 0.0171$ &
$13.337 \pm 0.062$ &
$0.182 \pm 0.026$ \\
DEB-only &
$8.38 \pm 0.78$ &
$+0.275 \pm 0.081$ &
$0.2967 \pm 0.0159$ &
\dots\tablenotemark{e} &
\dots\tablenotemark{e} \\
sCMD $\times$ DEB &
$8.46 \pm 0.66$ &
$+0.280 \pm 0.079$ &
$0.2968 \pm 0.0158$ &
$13.333 \pm 0.058$ &
$0.183 \pm 0.024$ \\
\enddata
\tablenotetext{a}{\citet{Brogaard2012} quote [Fe/H] with separate random and systematic components; here we combine them in quadrature for a single representative uncertainty.}
\tablenotetext{b}{\citet{Chaboyer1999} do not quote a $Y$ value.}
\tablenotetext{c}{\citet{VandenBerg_2014} give a true distance modulus, $(m-M)_0 = 13.05$, not an apparent $V$-band distance modulus.}
\tablenotetext{d}{\citet{McKeever2019} report an age and $Y$ from detailed frequency modeling; other quantities are model-dependent and are not consistently reported in the same way as isochrone-based cluster fits.}
\tablenotetext{e}{\textit{The DEB-only analysis constrains age, [Fe/H], and $Y$, but does not independently constrain $(m-M)_V$, or $E(B-V)$.}}
\end{deluxetable*}

\subsection{MC Error Budget}\label{subsec:error_budget}

\begin{figure}[!ht] 
  \centering
  \includegraphics[width=12cm]{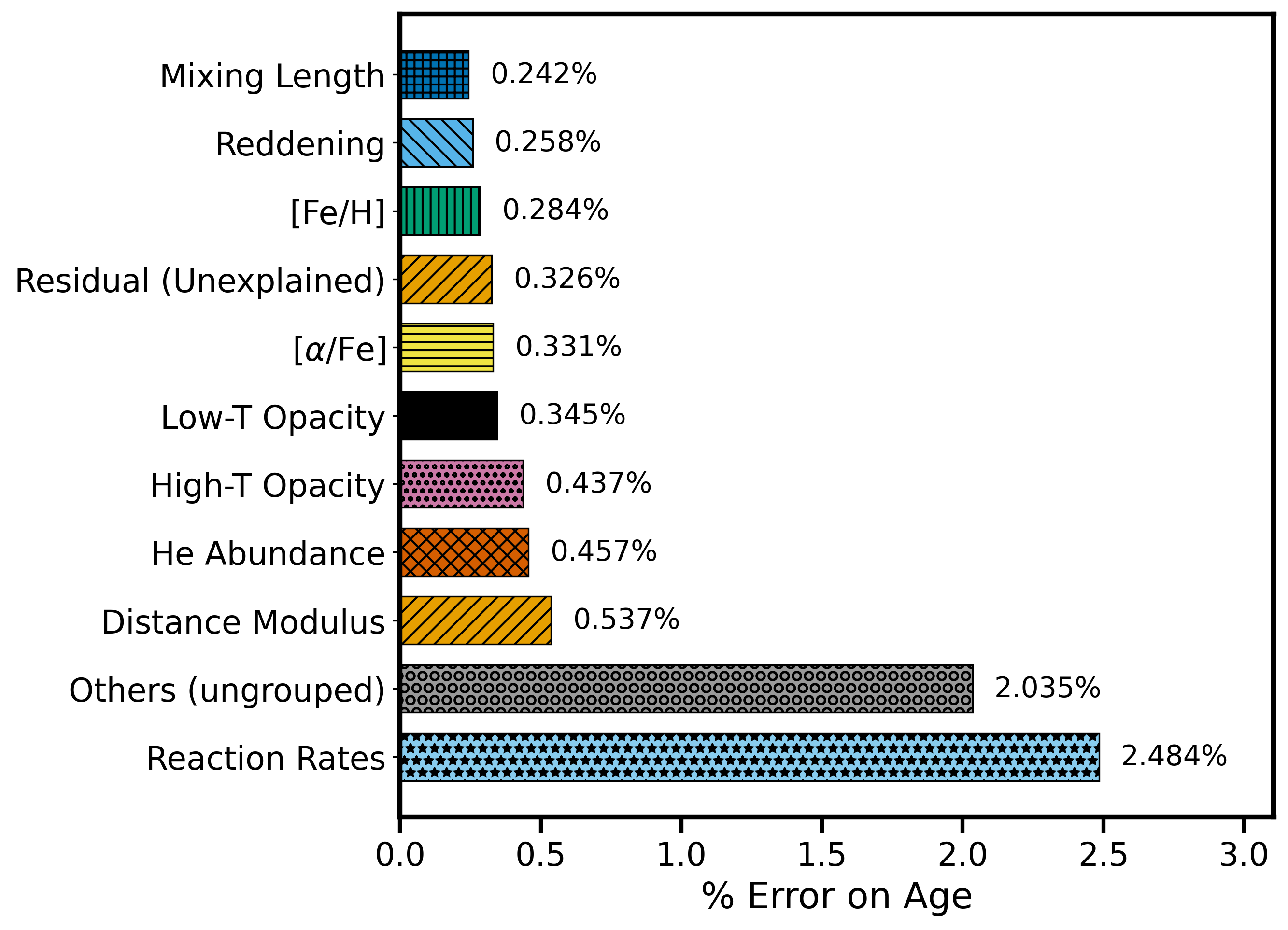}
  \caption{Contributions to the error of the age of NGC 6791 from each Monte Carlo parameter as well as distance and reddening. All the errors are converted to the percentage of error of the age of NGC 6791. The total age error is 7.74\%, of which the unexplained residual contributes about 0.33\%.}
  \label{fig:johnson_error_budget_stacked_like_ying}
\end{figure}

Our Monte Carlo framework yields marginalized posterior distributions for the cluster age and all nuisance parameters, enabling a direct decomposition of the age uncertainty into contributions from individual inputs. Following \citet{Ying2023,Ying2024,Ying2025}, we quantify the relative importance of each Monte Carlo parameter using the Johnson (2000) relative-weights method \citep{Johnson2000}, an efficient approximation to dominance analysis (LMG; \citealt{lindeman1980introduction}). In this approach, the correlated predictors are transformed into an orthogonal basis and the model $R^2$ is decomposed into contributions per-parameter, which we convert into fractional contributions to the age uncertainty. Figure~\ref{fig:johnson_error_budget_stacked_like_ying} summarizes the resulting error budget; the Other category groups parameters with individually small contributions.

For NGC\,6791, no single parameter dominates the age uncertainty. The error budget is broadly distributed across many inputs: the combined contributions from $[\mathrm{Fe/H}]$, He abundance, distance modulus, extinction, mixing length, and [$\alpha$/Fe] account for only about 28\% of the total, while the nuclear reaction-rate parameters together contribute about 34\%, with individual reactions such as $^3$He+$^3$He and $^3$He+$^4$He contributing about 8\% and 6\%, respectively. This contrasts with \citet{Ying2025}, who found distance and reddening to be the dominant sources of uncertainty in globular-cluster ages. The contrast emphasizes that error budgets can differ substantially between globular and open clusters and should be treated accordingly. In our case, reddening contributes less than five percent, which may reflect the fact that, unlike the two-filter photometry used by \citet{Ying2025}, our three-filter \emph{Gaia} photometry makes the reddening vector less degenerate with age-sensitive CMD morphology, especially around the main-sequence turnoff and subgiant branch. The contribution from [$\alpha$/Fe] is likewise modest. Although the PHOENIX atmosphere grid \citep{Husser2013} spans [$\alpha$/Fe] values from $-0.2$ to $+1.2$ in steps of 0.2 dex, our adopted prior, [$\alpha$/Fe] $= +0.06 \pm 0.08$, is sufficiently narrow that most realizations map to only a small subset of those tables: 76\% use the 0.0 table, 22\% use the $+0.2$ table, and 2\% use the $-0.2$ table, with a negligible fraction assigned to higher values. The relatively small uncertainty in [$\alpha$/Fe], together with the coarse spacing of the atmosphere grid, therefore limits its impact on the inferred age uncertainty.

\section{Discussion}\label{sec:discussion}

\subsection{NGC 6791 as a Galactic Archaeology Calibrator}\label{subsec:galactic_archeology}
\begin{figure*}[!ht] 
  \centering
  \includegraphics[width=0.59\textwidth]{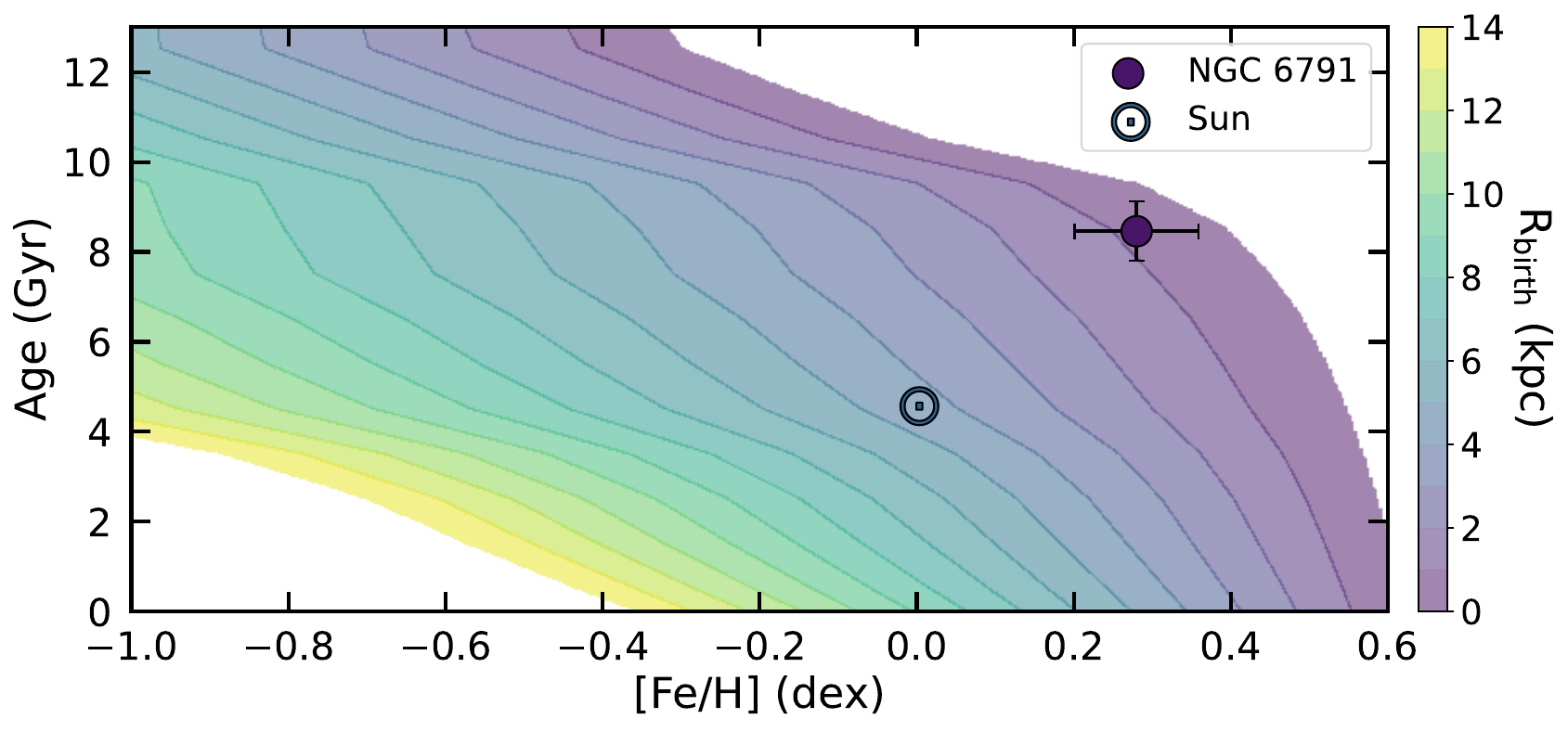}
    \includegraphics[width=0.38\textwidth]{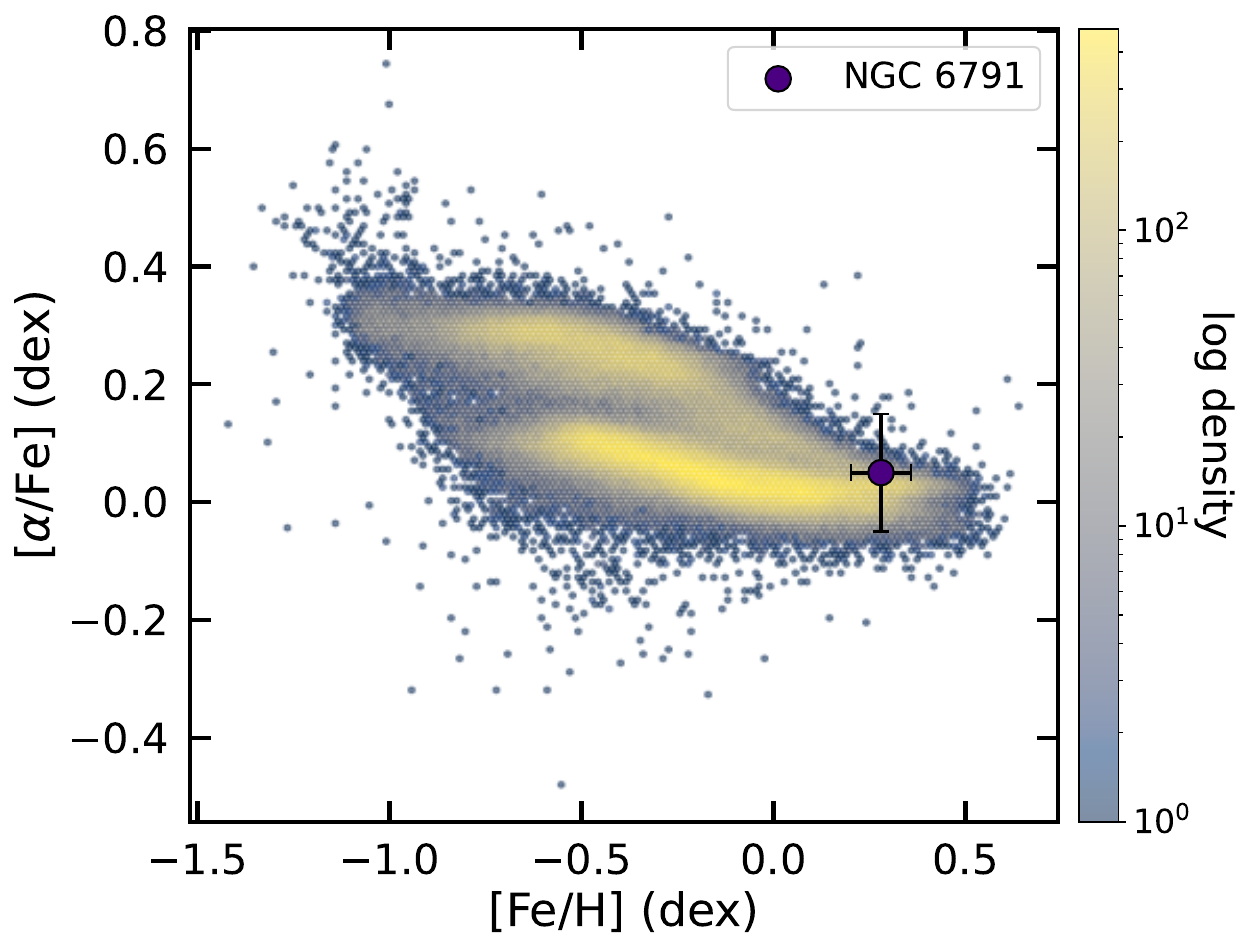}
  \caption{\textit{Left:} \feh-Age plane with NGC 6791 and Sun plotted and \rbirth\ contours shown in color using \rbirth\ calibrations from \cite{Lu2022}. From its newly measured age and \feh, NGC 6791 has an inferred \rbirth\ of 0 kpc, suggesting an inner-disk/bulge origin and a migration of $\sim 8$ kpc. \textit{Right:} \feh-\alphafe\ plane displaying high- and low-$\alpha$ sequences (thin/thick disks) from sample APOGEE data with NGC 6791 plotted and showing a membership at the tail end of the high-$\alpha$ sequence. }
  \label{fig:NGC6791Rbirth}
\end{figure*}

Our more accurate absolute age determination of $8.46^{+0.61}_{-0.52}$ Gyr for NGC 6791 allows a careful application of this cluster as a Galactic archaeology calibrator. In the left hand panel of Figure \ref{fig:NGC6791Rbirth}, we show its location on the \feh-age plane. These two stellar parameters allow stars to be traced back to their approximate Galactic birth radius, \rbirth, based on the radial metallicity gradient across the Galactic disk \citep[e.g.,][]{Minchev18,Frankel19,Ness19}. We show the \rbirth\ contours in color based on the empirical relation constructed in \cite{Lu2022}. The combination of an old age and super-solar metallicity (\feh= $0.30^{+0.10}_{-0.09}$) corresponds to an inferred birth radius of $\rbirth\ \lesssim 2$~kpc, consistent with theorized formation in the inner disk or bulge region followed by outward migration to its present-day location at 8 kpc \citep{Linden2017}. We emphasize that the inferred \rbirth\ is derived empirically and should be interpreted in a relative rather than absolute sense \citep{Lu2022b}. Nonetheless, the robust combination of NGC 6791's old age and super-solar metallicity provides a strong empirical constraint on the timescale of chemical evolution in the inner Milky Way.

Constraining this cluster's birth origin places an upper limit on when the high-$\alpha$ sequence stopped forming stars in the inner-disk. In the right hand panel of Figure \ref{fig:NGC6791Rbirth}, we show the high and low-$\alpha$ sequence of the Galactic disk from APOGEE survey data \citep{APOGEEDR17}. NGC~6791 lies near the low-$\alpha$ sequence for its metallicity, despite its old age and commonly assigned thick-disk kinematics (see Figure 20 from \citealt{Ahmed2025}). Its inferred \rbirth $\lesssim 2$~kpc and super-solar metallicity at $\sim 8.5$~Gyr imply that the inner Galaxy had already enriched to high metallicity by this epoch. This implies that the inner thick disk continued forming stars until at least $\sim 8$--9~Gyr ago and enriched to \feh\ $\gtrsim +0.3$ dex. This interpretation of our single object is consistent with the population-scale study of APOGEE stars for an independent CMD-based reconstruction of the Galactic disk star formation history by \cite{Fernandez25}. The thick-disk component is dominated by stars older than $\sim$10 Gyr indicating the bulk of its star formation occurred at early times, but with a non-negligible tail extending to younger ages ($\sim$4--6 Gyr, see Figure~7). The age of NGC~6791 places it within this extended tail, between the dominant early formation epoch and the youngest stars associated with this component. NGC~6791 could represent one of the latest and significant generations of thick-disk star formation in the inner Galaxy, providing an empirical anchor for the upper metallicity extent and timescale of thick-disk formation at small Galactocentric radii.


\subsection{NGC 6791 as an asteroseismic benchmark}\label{subsec:astro_v_iso}

NGC\,6791 is a useful calibrator for asteroseismic age work on RGB stars. Large surveys now provide global seismic ages for many giants \citep[e.g.,][]{Pinsonneault2025,Theodoridis2026}, but such ages benefit from benchmark clusters with well determined properties. NGC\,6791 is especially valuable in this context because it is old, super-solar in metallicity, lies in the \emph{Kepler} field, and hosts many oscillating giants. Its metal-rich regime is also important because simple scaling-relation inferences are known to be more challenging there, and previous studies have suggested that corrections to the seismic mass scale may be required for NGC\,6791-like stars \citep{Pinsonneault2018,Gaulme2016}.

\begin{figure}[!ht] 
  \centering
  \includegraphics[width=8cm]{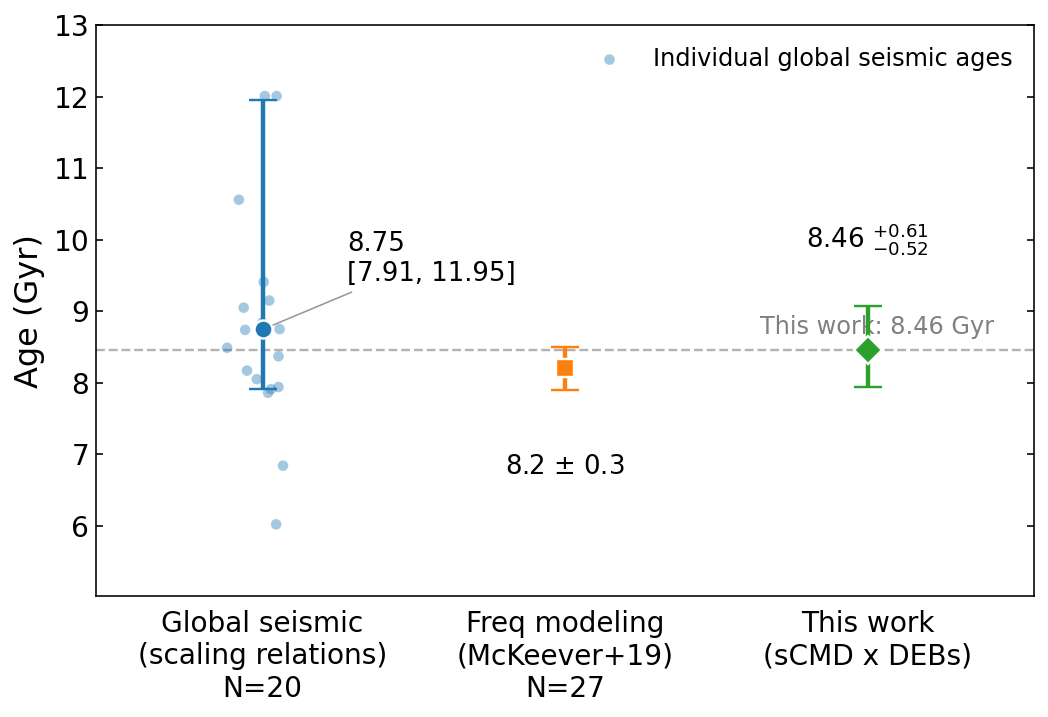}
    \caption{Asteroseismic age estimates for NGC\,6791 compared to benchmark constraints. 
    Left: individual global-parameter seismic ages (scaling relations; $N=20$) with the median and central 68\% interval annotated.
    Middle: age from detailed frequency modeling \citep{McKeever2019}.
    Right: our absolute age from combined sCMD and DEB constraints. The dashed line marks our central value.}  
    \label{fig:astro_v_iso}
\end{figure}

Figure~\ref{fig:astro_v_iso} compares several age estimates for NGC\,6791. The detailed frequency-modelling result agrees well with our independently derived cluster age \citep{McKeever2019}, while the global seismic ages of nominal cluster members show substantially larger star-to-star scatter \citep{Tayar2025,Pinsonneault2025}. We do not attempt to identify the origin of that dispersion here. Instead, our point is that the absolute age derived in this work provides an updated external benchmark in the super-solar metallicity regime. This benchmark complements existing calibrators at lower metallicity and can help anchor future asteroseismic age calibrations across a wider range of stellar populations. Extending absolute-age measurements to additional open and globular clusters in the \emph{Kepler} and \emph{K2} fields will be important for clarifying the origin of the observed cluster-to-cluster and star-to-star differences in seismic age estimates.

\subsection{Age Metallicity Relation of MW Clusters}\label{subsec:mw_clusters_age_metalicity}

\begin{figure}[!ht] 
  \centering
  \includegraphics[width=8cm]{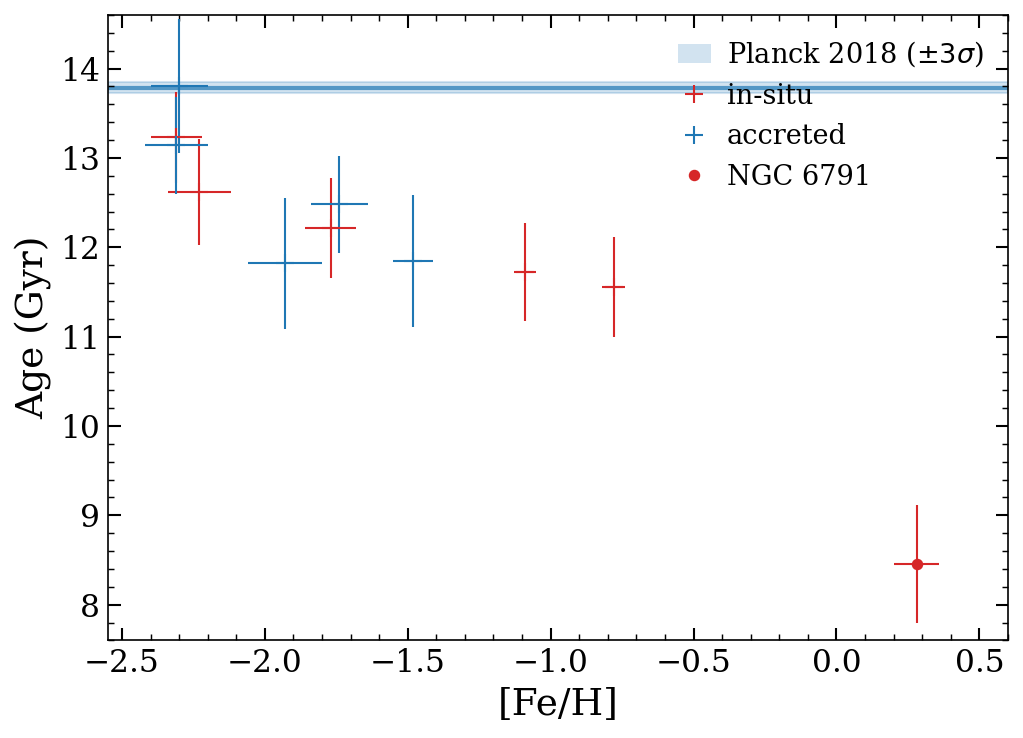}
  \caption{Absolute age vs metallicity plot for 10 Milky Way GCs and NGC\,6791. The age of the universe, as measured by \citet{PlanckCollaboration2020}, is also shown with its $3\sigma$ uncertainty.}
  \label{fig:Age_Metallicity}
\end{figure}

Combining our results with those of \citet{Ying2023,Ying2024,Ying2025}, we place 10 globular clusters and the open cluster NGC\,6791 in the absolute age–metallicity plane (Figure~\ref{fig:Age_Metallicity}). Section~\ref{subsec:galactic_archeology} used the age and metallicity of NGC\,6791 to interpret its likely birth environment, here we place the cluster in the broader Milky Way cluster population to examine how it extends the empirical age-metallicity relation across cluster types and Galactic regions. We adopt the in-situ versus accreted Milky Way globular cluster classifications from \citet{Belokurov2024}, which are based on total energy and angular momentum. Within this framework, NGC\,6791 is securely identified as an in-situ object, having formed within the Milky Way rather than being accreted from an external system. This interpretation is supported by the chemical and dynamical analysis of \citet{Seshashayana2024}, who show that NGC\,6791 shares the enrichment patterns of the inner disk and bulge populations rather than those of accreted satellites. NGC\,6791 thus provides an important link between globular and open clusters, occupying the young, metal-rich end of the age-metallicity plane while remaining broadly consistent with the negative correlation between absolute age and [Fe/H] observed in  Galactic globular clusters \citep{Ying2025}.  

Recent work on the age--metallicity relations of Milky Way globular clusters has emphasized that the GC age-metallicity relation (AMR) may contain progenitor-dependent structure, and that homogeneous age scales are essential for interpreting these sequences \citep{Massari2023,Massari2026,Lardo2026arXiv,Seshashayana2024}. We therefore do not use Figure~\ref{fig:Age_Metallicity} to infer the progenitor-dependent structure of the GC AMR or to draw conclusions about the separation between in-situ and accreted globular-cluster sequences. Instead, the figure is intended as an illustrative comparison placing NGC\,6791 on the same absolute age and metallicity scale as previously absolute age-dated globular clusters \citep{Ying2023,Ying2024,Ying2025}. Because young, metal-rich globular clusters are sparsely sampled, the current data do not allow a direct test of whether NGC\,6791 lies on an extrapolation of the in-situ globular-cluster AMR. Expanding the sample of clusters with homogeneous absolute ages, particularly young globular clusters and old open clusters at high metallicity, will be necessary to determine whether NGC\,6791 represents a continuous extension of the in-situ cluster AMR or instead traces a distinct old open-cluster sequence associated with the chemically enriched inner Galaxy. Such an expanded absolute-age catalog would also provide a stronger empirical basis for future studies of globular-cluster assembly and the early formation history of the Milky Way.



\section{Conclusion}\label{sec:conclusion}

We derive an absolute age for NGC\,6791 from Gaia photometry using a combination of synthetic color–magnitude diagrams (sCMDs) and detached eclipsing binaries (DEBs), evaluated against Monte Carlo isochrone realizations that marginalize over distance, reddening and uncertainties in the physics used in stellar evolution models. We find the  age  to be $8.46 \pm 0.66~\mathrm{Gyr}$, a metallicity of $[\mathrm{Fe/H}] = +0.280 \pm 0.079$, an initial helium abundance of $Y = 0.2968 \pm 0.0158$, an apparent distance modulus of $(m-M)_V = 13.333 \pm 0.058$, and a reddening of $E(B-V) = 0.183 \pm 0.024$.

Our Monte Carlo framework marginalizes over key stellar-physics inputs in the Dartmouth Stellar Evolution Program, including convection and transport prescriptions, microphysics, and nuclear reaction rates, while also varying composition and external parameters. To model the observed CMD broadening, we inject empirical scatter using a seam-carving approach with a normalized distance metric, enabling applications to clusters lacking dedicated artificial-star tests.  We combine two complementary constraints through bootstrap-calibrated weights: the CMD morphology, quantified with a 2D KS statistic, and the DEB constraints, quantified with a nearest-point $\chi^2$ statistic in $(M,L)$ space.

Our inferred parameters broadly agree with literature determinations obtained using a range of independent methods. Small differences in distance modulus and reddening are expected given differing membership selections and the use of Gaia photometry in this work. The age uncertainty in NGC 6791 is not dominated by any single source, but is instead distributed across many inputs. The combined contributions from $[\mathrm{Fe/H}]$, He abundance, distance modulus, extinction, mixing length, and [$\alpha$/Fe] account for only about 28\% of the total age uncertainty, while nuclear reaction-rate parameters together contribute about 34\%, with individual reactions such as $^3$He+$^3$He and $^3$He+$^4$He contributing about 8\% and 6\%, respectively  underscoring their importance in stellar evolution modeling. We find that the error budget differs substantially between globular clusters and open clusters and should be treated accordingly. 

With its super-solar metallicity, NGC\,6791 provides a stringent constraint on early chemical enrichment in the inner Galaxy and on radial migration scenarios from its center. Our results also extend the cluster absolute age–metallicity relation in the Milky Way and reinforce the role of NGC\,6791 as a key benchmark for comparing global asteroseismology with more local seismic diagnostics. Future work will expand this absolute-age framework to additional clusters to further populate the Milky Way cluster age–metallicity plane and to strengthen the network of benchmark systems used in modern asteroseismic calibration.

\begin{acknowledgments}
We thank the anonymous referee whose constructive report improved the presentation of this paper. We thank E. Boudreaux and J. M. Ying for their useful discussions throughout this work. 
\end{acknowledgments}

\section*{Data Availability}

The calibrated HST/ACS photometry for NGC~6791 used in this work was retrieved from MAST as part of the HST UV Globular-Cluster Survey \citep[HUGS;][]{hugs}.

\software{\texttt{isochronetoolbox} \citep{isochronetoolbox_github_2026}, \texttt{Fidanka} \citep{Boudreaux_fidanka_2023}}

\appendix

\section{Electronic Photometry Table}
\label{appendix:photometry}

Table~\ref{tab:photometry_sample} presents a sample of the cleaned Gaia DR3 photometry for NGC~6791 stars with $15.8 \leq G \leq 18.8$ mag. The full table is available electronically.

\begin{table*}[ht]
\centering
\caption{Sample of cleaned Gaia DR3 photometry for NGC~6791 in the range $15.8 \leq G \leq 18.8$ mag. The full table is available electronically.}
\label{tab:photometry_sample}
\begin{tabular}{cccccc}
\hline
star\_id & original\_star\_id & BP\_RP & Gmag & ra & dec \\
\hline
0 & 2.05136E+18 & 1.191127 & 18.823915 & 289.8320228 & 38.23505004 \\
1 & 2.05137E+18 & 1.066367 & 17.939232 & 289.8816805 & 38.24299786 \\
2 & 2.05138E+18 & 1.097950 & 17.308966 & 289.6286292 & 38.30633379 \\
3 & 2.05133E+18 & 1.147731 & 17.188213 & 289.4625129 & 37.91248849 \\
4 & 2.05135E+18 & 1.075452 & 17.903330 & 289.7131044 & 37.96323188 \\
5 & 2.05136E+18 & 1.101250 & 17.985266 & 289.7430462 & 38.03602389 \\
\hline
\end{tabular}
\end{table*}

\bibliographystyle{aasjournalv7}
\bibliography{sample701}


\end{document}